\documentclass{ws-p10x7}

\newcommand{\gsim}{\mathrel{\raisebox{-.6ex}{$\stackrel{\textstyle>}{\sim}$}}}

\begin{document}
\flushbottom

\title{THE HIGHEST ENERGY COSMIC RAYS, GAMMA-RAYS AND NEUTRINOS:
FACTS, FANCY and RESOLUTION}

\author{FRANCIS HALZEN}

\address{Department of Physics, University of Wisconsin, Madison, WI
53706, USA}

\twocolumn[\maketitle\abstract{
Although cosmic rays were discovered 90 years ago, we do not know how and
where they are accelerated. There is compelling evidence that the
highest energy cosmic rays are extra-galactic --- they cannot be
contained by our galaxy's magnetic field anyway because their
gyroradius exceeds its dimensions. Elementary elementary-particle
physics dictates a universal upper limit on their energy of
$5\times10^{19}$\,eV, the so-called Greisen-Kuzmin-Zatsepin cutoff;
however, particles in excess of this energy have been observed, adding one
more puzzle to the cosmic ray mystery. Mystery is nonetheless  fertile ground
for progress: we will review the facts and mention some  very
speculative interpretations. There is indeed a realistic hope that
the oldest problem in astronomy will be resolved soon by ambitious
experimentation: air shower arrays of $10^4$\,km$^2$ area,
arrays of air Cerenkov detectors and kilometer-scale neutrino
observatories.}]

\section{The New Astronomy}

Conventional astronomy spans 60 octaves in photon frequency, from
$10^4$\,cm radio-waves to $10^{-14}$\,cm photons of GeV energy; see
Fig.\,1. This is an amazing expansion of the power of our eyes which
scan the sky over less than a single octave just above $10^{-5}$\,cm
wavelength. The new astronomy, discussed in this talk, probes the
Universe with new wavelengths, smaller than $10^{-14}$\,cm, or photon
energies larger than 10\,GeV. Besides gamma rays, gravitational waves
and neutrinos as well as very high energy protons that are only
weakly deflected by the
magnetic field of our galaxy, become astronomical messengers from the
Universe. As exemplified time and again, the development of novel
ways of looking into space invariably results in the discovery of
unanticipated phenomena. As is the case with new accelerators,
observing the predicted will be slightly disappointing.

\begin{figure*}[t]
\centering\leavevmode
\epsfxsize=9cm
\epsffile{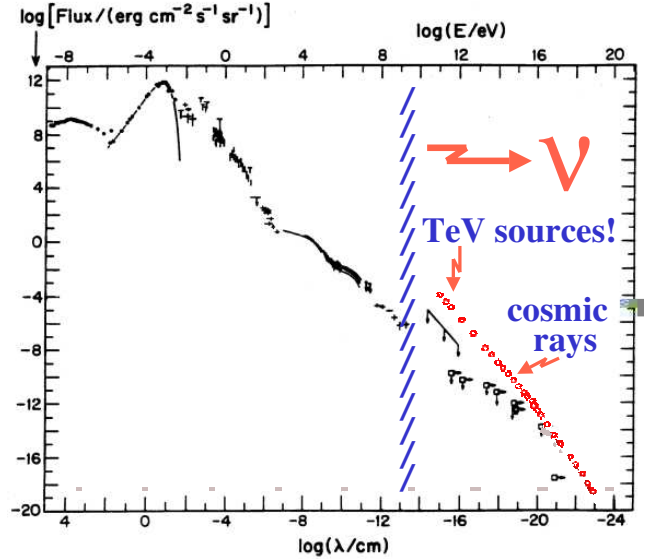}
\caption{The diffuse flux of photons in the Universe, from radio waves to GeV-photons. Above tens of GeV only limits are reported although individual sources emitting TeV gamma-rays have been identified. Above GeV energy cosmic rays dominate the spectrum.}
\end{figure*}

Why do high energy astronomy with neutrinos or protons despite the considerable instrumental challenges which we will discuss further on? A mundane reason is that the
Universe is not transparent to photons of TeV energy and above
(units are: GeV/TeV/PeV/EeV/ZeV in ascending factors of $10^3$). For
instance, a PeV energy photon $\gamma$ cannot reach us from a source
at the edge of our own galaxy because it will annihilate into an
electron pair in an encounter with a 2.7 degree Kelvin microwave
photon $\gamma_{\rm CMB}$ before reaching our telescope. Energetic
photons are absorbed on background light by pair production $\gamma +
\gamma_{\rm \,bkgnd} \rightarrow e^+€ + e^-$ of electrons above a
threshold $E$ given by
\begin{equation}
4E\epsilon \sim (2m_e)^2 \,,
\end{equation}
where $E$ and $\epsilon$ are the energy of the high-energy and
background photon, respectively. Eq.\,(1) implies that TeV-photons are absorbed
on infrared light, PeV photons on the cosmic microwave background and
EeV photons on radio-waves. Only neutrinos can reach us without
attenuation from the edge of the Universe.

At EeV energies proton astronomy may be possible. Near 50\,EeV and
above, the arrival directions of electrically charged cosmic rays are
no longer scrambled by the ambient magnetic field of our own galaxy.
They point back to their sources with an accuracy determined by their
gyroradius in the intergalactic magnetic field $B$:
\begin{equation}
\theta \cong {d\over R_{\rm gyro}} = {dB\over E} \,,
\end{equation}
where $d$ is the distance to the source. Scaled to units relevant to
the problem,
\begin{equation}
{\theta\over0.1^\circ} \cong { \left( d\over 1{\rm\ Mpc} \right)
\left( B\over 10^{-9}{\rm\,G} \right) \over \left( E\over
3\times10^{20}\rm\, eV\right) }\,.
\end{equation}
Speculations on the strength of the inter-galactic magnetic field
range from $10^{-7}$ to $10^{-12}$~Gauss. For a distance of 100~Mpc,
the resolution may therefore be anywhere from sub-degree to
nonexistent. It is still reasonable to expect that the arrival
directions of the highest energy cosmic rays provide
information on the location of their sources. Proton astronomy should
be possible; it may also provide indirect information on
intergalactic magnetic fields. Determining their strength by conventional astronomical means has turned out to be challenging.

\section{The Highest Energy Cosmic Rays: Facts}

In October 1991, the Fly's Eye cosmic ray detector recorded an event
of energy $3.0\pm^{0.36}_{0.54}\times 10^{20}$\,eV.\cite{flyes}
This event, together with an event recorded by the Yakutsk air shower
array in May 1989,\cite{yakutsk} of estimated energy $\sim$
$2\times10^{20}$\,eV, constituted at the time the two highest
energy cosmic rays ever seen. Their energy corresponds to a center of
mass energy of the order of 700~TeV or $\sim 50$ Joules, almost 50
times LHC energy. In fact, all experiments\cite{web} have detected
cosmic rays in the vicinity of 100~EeV since their discovery by the
Haverah Park air shower array.\cite{WatsonZas} The AGASA air shower
array in Japan\cite{agasa} has by now accumulated an impressive 10
events with energy in excess of $10^{20}$\,eV.\cite{ICRC}

\begin{figure*}[t]
\centering\leavevmode
\epsfxsize=11cm
\epsffile{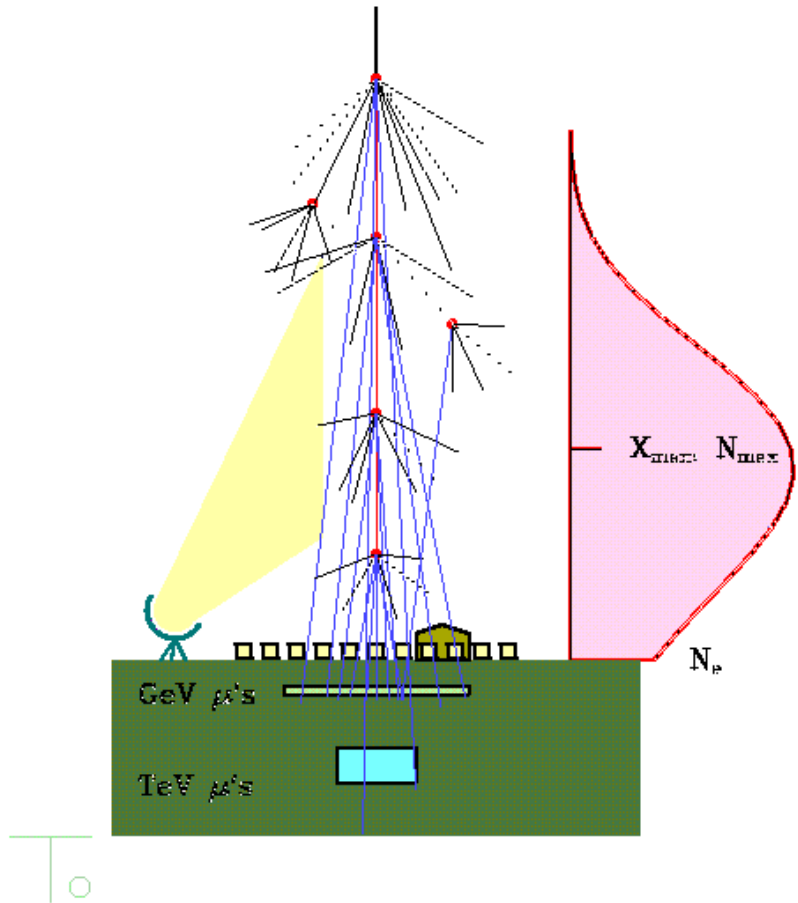}
\caption{Particles interacting near the top of the atmosphere initiate an electromagnetic and hadronic particle cascade. Its profile is shown on the right. The different detection methods are illustrated. Mirrors collect the Cerenkov and nitrogen fluorescent light, arrays of detectors sample the shower reaching the ground, and underground detectors identify the muon component of the shower.}
\end{figure*}

How well experiments can determine the energy of these events is a
critical issue. With a particle flux of order 1 event per km$^2$ per
century, these events can only be studied by using the earth's
atmosphere as a particle detector. The experimental signatures of a
shower initiated by a cosmic particle are illustrated in the cartoon
shown in Fig.\,2. The primary particle creates an electromagnetic and
hadronic cascade. The electromagnetic shower
grows to a shower maximum, and is subsequently absorbed by the
atmosphere. This leads to the characteristic shower profile shown on
the right hand side of the figure. The shower can be observed by: i)
sampling the electromagnetic and hadronic components when they reach€
the ground
with an array of particle detectors such as scintillators, ii)
detecting the fluorescent
light emitted by atmospheric nitrogen excited by the passage of the
shower particles, iii) detecting the Cerenkov light emitted by the€
large number
of particles at shower maximum, and iv)~detecting muons and neutrinos
underground. Fluorescent and Cerenkov light is collected by large
mirrors and recorded by arrays of photomultipliers in their focus. The
bottom line on energy measurement is that, at this time, several
experiments using the first two techniques agree on the energy of EeV-showers within a typical resolution of 25\%. Additionally, there is a systematic error of order 10\% associated with the modeling of the showers. All techniques are indeed subject to the ambiguity of particle simulations that involve physics beyond LHC. If the final outcome turns out to be erroneous inference of the energy of the shower because of new physics associated with particle interactions, we will be happy to contemplate this discovery instead.

\looseness=-1
Whether the error in the energy measurement could be significantly
larger is a key question to which the answer is almost certainly
negative. A variety of techniques have been developed to overcome the
fact that conventional air shower arrays do calorimetry by sampling
at a single depth. They give results within the range already
mentioned. So do the fluorescence experiments that embody continuous
sampling calorimetry. The latter are subject to understanding the
transmission of fluorescent light in the dark night atmosphere --- a
challenging problem given its variation with weather. Stereo
fluorescence detectors will eliminate this last hurdle by doing two
redundant
measurements of the same shower from different locations. The HiRes collaborators have one year of data on tape which should allow them to
settle any doubts as to energy calibration once and for all.

The premier experiments, HiRes and AGASA, agree that cosmic rays with
energy in excess of 10\,EeV are not a feature of our galaxy and that
their spectrum extends beyond 100\,EeV. They disagree on almost
everything else. The AGASA experiment claims evidence that they come
from point sources, and
that they are mostly heavy nuclei. The HiRes data do not support
this. Because of statistics, interpreting the measured fluxes as a
function of energy is like reading tea leaves; one cannot help however
reading different messages in the spectra (see Fig.\,3). More about
that later.

\begin{figure}[t!]
\epsfxsize=6.8cm
\epsffile{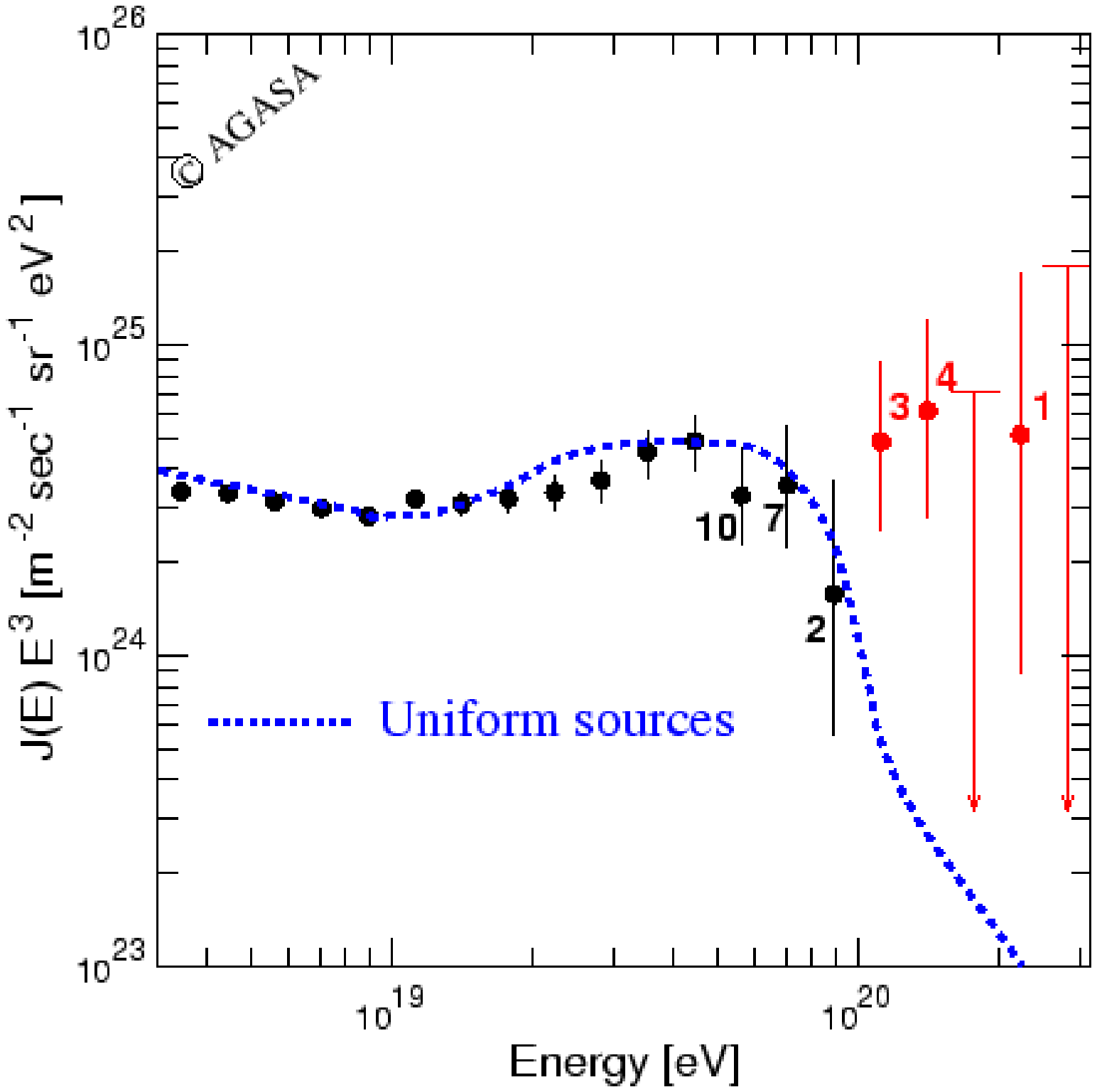}
\vspace{3ex}

\epsfxsize=6.8cm
\epsffile{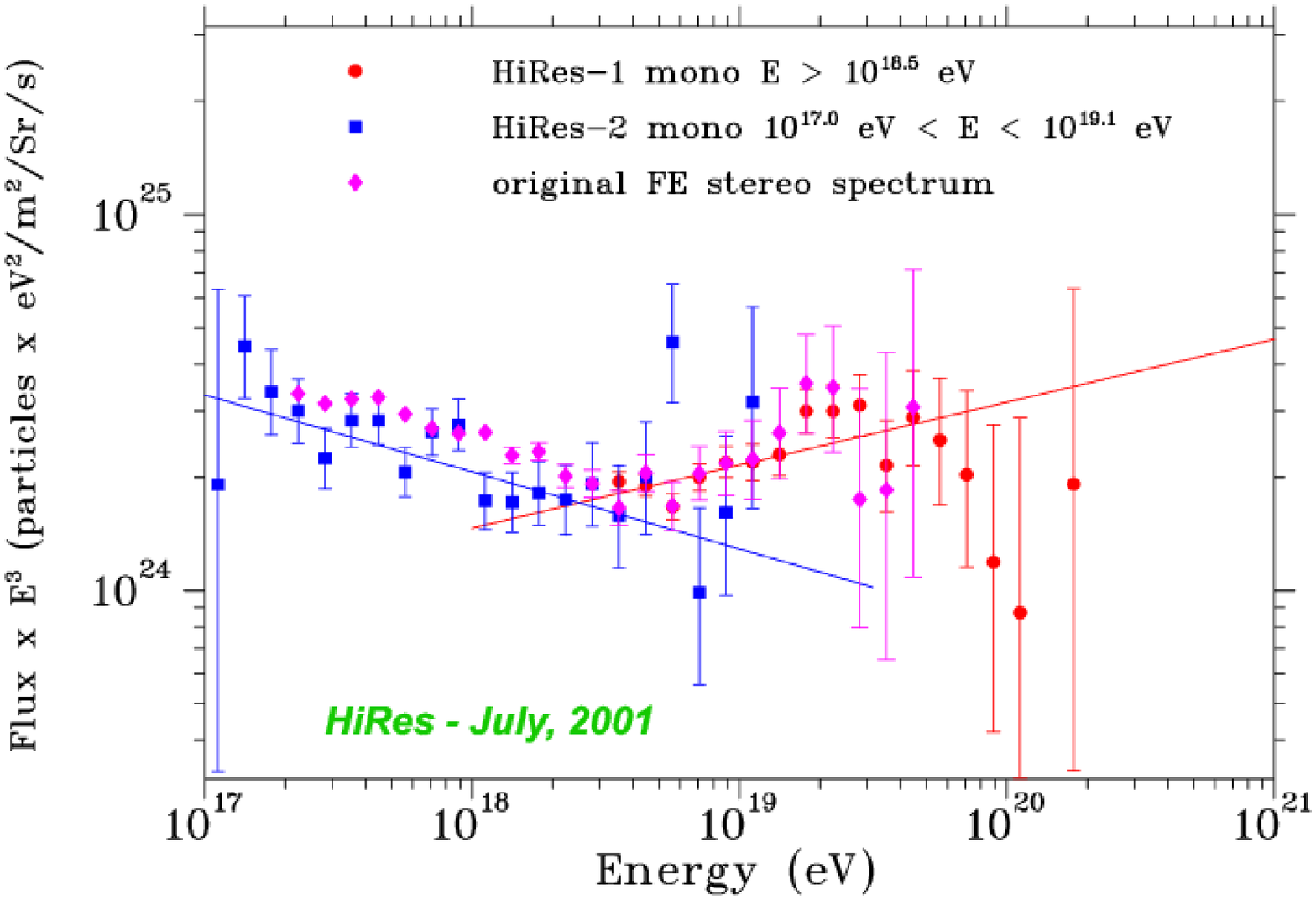}
\caption{The cosmic ray spectrum peaks in the vicinity of 1\,GeV and has features near $10^{15}$ and $10^{19}$\,eV. They are referred to as the ``knee" and ``ankle" in the spectrum. Shown is the flux of the highest energy cosmic rays near and beyond the ankle measured by the AGASA and HiRes experiments.}
\end{figure}

\section{The Highest Energy Cosmic Rays: Fancy}

\subsection{Acceleration to $>100$ EeV?}

\begin{table*}[t]
\centering\leavevmode
\begin{tabular}{|llll|}
\hline
\multicolumn{4}{|c|}{Conditions with $E \sim 10\rm\ EeV$}\\
\hline
$\bullet$\ quasars& $\Gamma\cong 1$& $B\cong 10^3$ G& $M\cong 10^9
M_{\rm sun}$\\
$\bullet$\ blasars& $\Gamma\gsim 10$& $B\cong 10^3$ G& $M\cong 10^9
M_{\rm sun}$\\
$\bullet$\ \parbox[t]{.8in}{neutron stars\\ black holes\\ $\vdots$}&
$\Gamma\cong 1$& $B\cong 10^{12}$ G& $M\cong M_{\rm sun}$\\
$\bullet$\ grb& $\Gamma\gsim 10^2$& $B\cong 10^{12}$ G& $M\cong M_{\rm sun}$\\
\hline
\end{tabular}
\end{table*}

It is sensible to assume that, in order to accelerate a proton to
energy $E$ in a magnetic field $B$, the size $R$ of the accelerator
must be larger than the gyroradius of the particle:
\begin{equation}
R > R_{\rm gyro} = {E\over B}\,.
\end{equation}
I.e. the accelerating magnetic field must contain the particle orbit. This condition yields a maximum energy
\begin{equation}
E = \Gamma BR
\end{equation}
by dimensional analysis and nothing more. The $\Gamma$-factor has
been included to allow for the possibility that we may not be at rest
in the frame of the cosmic accelerator resulting in the observation of boosted particle energies.
Theorists' imagination regarding the accelerators is
limited to dense regions where exceptional gravitational forces
create relativistic particle flows: the dense cores of exploding
stars, inflows on supermassive black holes at the centers of active
galaxies, annihilating black holes or neutron stars? All
speculations involve collapsed objects and we can therefore replace $R$
by the Schwartzschild radius
\begin{equation}
R \sim GM/c^2
\end{equation}
to obtain
\begin{equation}
E \sim \Gamma BM \,.
\end{equation}
Given the microgauss magnetic field of our galaxy, no structures are
large or massive enough to reach the energies of the highest energy
cosmic rays. Dimensional analysis therefore limits their sources to
extragalactic objects; a few common speculations are listed in
Table\,1. Nearby active galactic nuclei distant by $\sim100$~Mpc and
powered by a billion solar mass black holes are candidates.
With kilo-Gauss fields we reach 100\,EeV. The jets (blazars) emitted by the
central black hole could reach similar energies in accelerating
substructures boosted in our direction by a $\Gamma$-factor of 10,
possibly higher. The neutron star or black hole remnant of a
collapsing
supermassive star could support magnetic fields of $10^{12}$\,Gauss,
possibly larger. Shocks with $\Gamma > 10^2$ emanating from the
collapsed black hole could be the origin of gamma ray bursts and,
possibly, the source of the highest energy cosmic rays.

The above speculations are reinforced by the fact that the sources
listed happen to also be the sources of the highest energy gamma rays
observed. At this point however a
reality check is in order. Let me first point out that the above
dimensional analysis applies to the Fermilab accelerator: 10\,kGauss
fields over several kilometers yield 1\,TeV. The argument holds
because, with optimized design and perfect alignment of magnets, the
accelerator
reaches efficiencies matching the dimensional limit. It is highly
questionable that Nature can achieve this feat. Theorists can
imagine acceleration in shocks with efficiency of perhaps 10\%.

The astrophysics problem is so daunting that many believe that
cosmic rays are not the beam of cosmic accelerators but the decay
products of remnants from the early Universe, for instance topological
defects associated with a grand unified GUT phase transition. A topological defect
will suffer a chain decay into GUT particles X,Y, that subsequently
decay to familiar weak bosons, leptons and quark- or gluon jets. Cosmic
rays are the fragmentation products of these jets. We know from
accelerator studies that, among the fragmentation products of jets,
neutral pions (decaying into photons) dominate protons by two orders
of magnitude. Therefore, if the decay of topological defects is the source of the
highest energy cosmic rays, they must be photons. This is a problem
because the highest energy event observed by the Fly's Eye is not
likely to be a photon.\cite{vazquez} A photon of 300\,EeV will
interact with the
magnetic field of the earth far above the atmosphere and disintegrate
into lower energy cascades --- roughly ten at this particular energy.
The measured shower profile of the event does not support this
assumption; see Fig.\,4. One can live and die by a single event!

\begin{figure}[h]
\centering\leavevmode
\epsfxsize=7cm
\epsffile{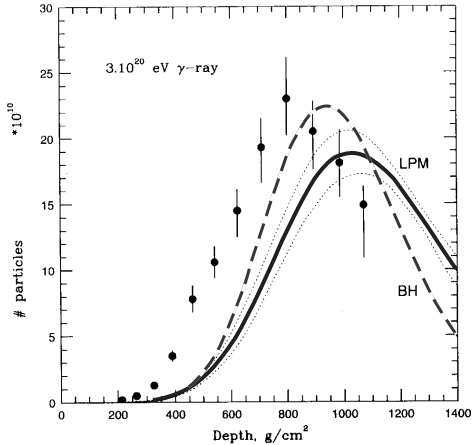}

\caption[]{
The composite atmospheric shower profile of a $3\times 10^{20}$\,eV
$\gamma$-ray shower calculated with Landau-Pomeranchuk-Migdal (solid) and Bethe-Heitler (dashed)
electromagnetic cross sections. The central line shows the average shower
profile and the upper and lower lines show 1~$\sigma$ deviations --- not
visible for the BH case, where lines overlap. The experimental shower profile
is shown along with the data points. It does not fit the profile of a photon shower.}
\end{figure}

\subsection{Are Cosmic Rays Really Protons: the GZK Cutoff?}

All experimental signatures agree on the particle nature of the
cosmic rays --- they look like protons, or, possibly, nuclei. We
mentioned at the beginning of this article that the Universe is
opaque to photons with energy in excess of tens of TeV because they annihilate into electron pairs in interactions with background light.
Also protons interact with background
light, predominantly by photoproduction of the $\Delta$-resonance,
i.e.\ $p + \gamma_{CMB} \rightarrow \Delta \rightarrow \pi + p$ above
a threshold energy $E_p$ of about 50\,EeV given by:
\begin{equation}
2E_p\epsilon > \left(m_\Delta^2 - m_p^2\right) \,.
\label{eq:threshold}
\end{equation}
The major source of proton energy loss is photoproduction of pions on
a target of cosmic microwave photons of energy $\epsilon$. The
Universe is therefore also opaque to the highest energy cosmic rays,
with an absorption length:
\begin{eqnarray}
\lambda_{\gamma p} &=& (n_{\rm CMB} \, \sigma_{p+\gamma_{\rm CMB}})^{-1}\\
&\cong& 10\rm Mpc,
\end{eqnarray}
or only tens of megaparsecs when their energy exceeds 50\,EeV. This
so-called GZK cutoff establishes a universal upper limit on
the energy of the cosmic rays. The cutoff is robust,
depending only on two known numbers: $n_{\mathrm{CMB}} = 400\rm\,cm^{-3}$ and
$\sigma_{p+\gamma_{\mathrm{CMB}}} = 10^{-28}\rm\,cm^2$.

Protons with energy in excess of 100\,EeV, emitted in distant quasars
and gamma ray bursts, will have lost their energy to pions before
reaching our detectors. They have, nevertheless, been observed, as we have previously discussed. They do not point to any sources within the GZK-horizon however, i.e. to sources in our local cluster of galaxies. There are three possible resolutions: i) the protons are accelerated in nearby sources, ii)~they do reach us from distant sources which accelerate them to much higher energies than we observe, thus exacerbating the acceleration problem, or iii) the highest energy cosmic rays are not protons.

The first possibility raises the challenge of finding an appropriate
accelerator by confining these already unimaginable sources to our
local galaxy cluster. It is not impossible that all cosmic rays are
produced by the active galaxy M87, or by a nearby gamma ray burst
which exploded a few hundred years ago. The sources identified by the AGASA array do not correlate however with any such candidates. 

Stecker\cite{stecker2} has speculated that the highest energy cosmic 
rays are Fe nuclei with a delayed GZK cutoff. The details are 
compicated but the relevant quantity in the problem is $\gamma=E/AM$, 
where A is the atomic number and M the nucleon mass. For a fixed 
observed energy, the smallest boost above GZK threshold is associated 
with the largest atomic mass, i.e.~Fe.

\subsection{Could Cosmic Rays be Photons or Neutrinos?}

When discussing topological defects, I already challenged the possibility that the original Fly's Eye event is a photon. The detector collects light produced by the fluorescence
of atmospheric nitrogen along the path of the high-energy shower
traversing
the atmosphere. The anticipated shower profile of a 300\,EeV photon
is shown in Fig.\,4. It disagrees with the data.

The observed shower profile roughly fits that of a primary proton, or, possibly, that of a nucleus. The shower profile information is however
sufficient to conclude that the
event is unlikely to be of photon origin. The same conclusion is
reached for the Yakutsk event that is characterized by a huge number
of secondary muons, inconsistent with an electromagnetic cascade
initiated by a gamma-ray. Finally, the AGASA collaboration claims evidence
for ``point" sources above 10\,EeV. The arrival directions are
however smeared out in a way consistent with primaries deflected by
the galactic magnetic field. Again, this indicates charged primaries
and excludes photons.

Neutrino primaries are definitely ruled out. Standard model neutrino
physics is understood, even for EeV energy. The average $x$ of
the parton mediating the neutrino interaction is of
order $x \sim \sqrt{M_W^2/s} \sim 10^{-6}$ so that the perturbative
result for the neutrino-nucleus cross section is calculable from measured HERA structure functions. Even at 100\,EeV a
reliable value of the cross section can be obtained based on QCD-inspired extrapolations of the structure function. The neutrino cross section is known to better than an order of magnitude. It falls 5 orders of magnitude short of the strong cross sections required to make a neutrino interact in the upper atmosphere to create an air shower.

Could EeV neutrinos be strongly interacting because of new physics?
In theories with TeV-scale gravity one can imagine that graviton
exchange dominates all interactions and thus erases the difference between quarks and
neutrinos at the energies under consideration. Notice however that
the actual models performing this feat require a fast turn-on of the cross
section with energy that violates S-wave unitarity.\cite{han}

We thus exhausted the possibilities: neutrons, muons and other
candidate primaries one may think of are unstable. EeV neutrons
barely live long enough to reach us from sources at the edge of our galaxy.

\section{A Three Prong Assault on the Cosmic Ray Puzzle}

We conclude that, where the highest energy cosmic rays are concerned, both the
accelerator mechanism and the particle physics are totally
enigmatic. The mystery has inspired a worldwide effort to tackle the
problem with novel experimentation in three complementary areas of
research: air shower detection, atmospheric Cerenkov astronomy and
underground neutrino physics. While some of the future instruments
have other missions, all are likely to have a major impact on cosmic
ray physics.

\subsection{Giant Cosmic Ray Detectors}

With super-GZK fluxes of the order of a single event per
kilometer-squared per century, the outstanding problem is the lack of
statistics; see Fig.\,3. In the next five years, a qualitative improvement can be expected from the operation of the HiRes fluorescence detector in Utah.
With improved instrumentation yielding high quality
data from 2 detectors operated in coincidence, the interplay between
sky transparency and energy
measurement can be studied in detail. We can safely anticipate that
the existence of super-Greisen energies will be conclusively
demonstrated by using the instrument's calorimetric
measurements.€ A mostly Japanese collaboration has proposed a
next-generation fluorescence detector, the Telescope Array.

The Auger air shower array is tackling the low rate problem with a huge
collection area covering 3000 square kilometers on an elevated plain
in Western Argentina. The instrumentation consists of 1600 water
Cerenkov detectors spaced by 1.5\,km. For calibration, about 15
percent of the showers occurring at night will be viewed by 3
HiRes-style fluorescence detectors. The detector will observe several
thousand events per year above 10\,EeV and tens above 100\,EeV, with
the exact numbers depending on the detailed shape of the observed
spectrum which is at present a matter of speculation; see Fig.\,3.

\subsection{Gamma-Rays from Cosmic Accelerators}

\begin{figure*}[t]
\centering\leavevmode
\epsfxsize=9cm
\epsffile{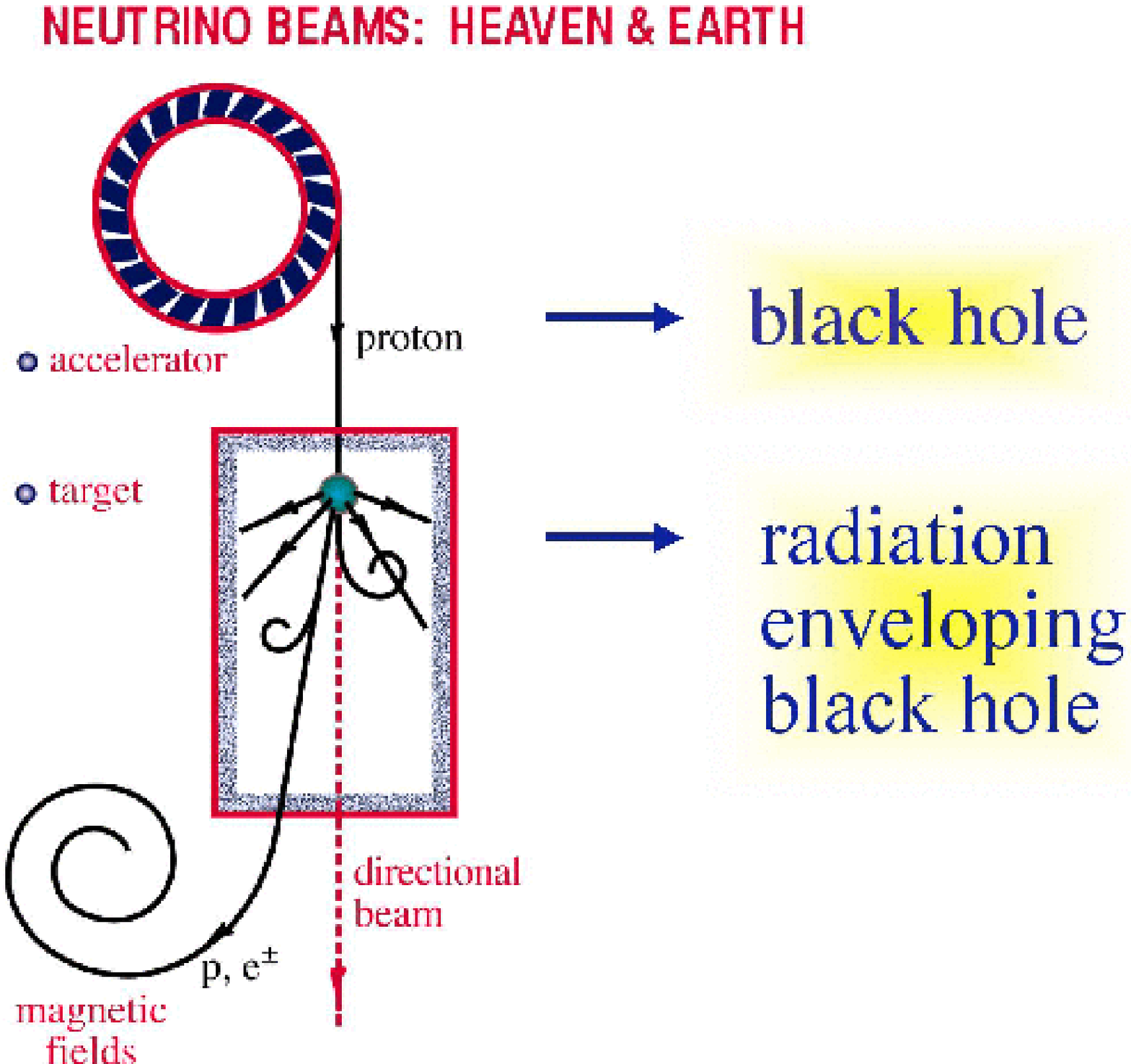}
\caption{}
\end{figure*}

An alternative way to identify the sources of the cosmic rays is
illustrated in Fig.\,5. The cartoon draws our attention to the fact that cosmic accelerators are also cosmic beam dumps producing secondary photon and neutrino beams.
Accelerating particles to TeV energy and above requires high-speed,
massive bulk flows.€ These are likely to have their origin in
exceptional gravitational forces associated with dense
cores of exploding stars, inflows onto supermassive black holes
at the centers of active galaxies, annihilating black holes or
neutron stars.€ In such situations, accelerated particles
are likely to pass through intense radiation fields or dense
clouds of gas leading to production of secondary
photons and neutrinos that accompany the primary cosmic-ray beam.
An example of an electromagnetic beam dump is the X-ray
radiation fields surrounding the central black holes of active galaxies.
The target material, whether a gas or particles or of photons,
is likely to be sufficiently tenuous so that the primary beam
and the photon beam are only partially attenuated.€ However,
it is also a real possibility that one could have a shrouded
source from which only the neutrinos can emerge, as in 
terrestrial beam dumps at CERN and Fermilab.

The astronomy€ event of the 21st century could be the simultaneous
observation of TeV-gamma rays, neutrinos and gravitational waves from
cataclysmic events associated with the source of the cosmic rays.

We first concentrate on the possibility of detecting high-energy
photon beams. After two decades, ground-based gamma ray astronomy has
become a mature science.\cite{weekes} A large mirror, viewed by an
array of photomultipliers, collects the Cerenkov light emitted by air
showers and images the showers in order to determine the arrival
direction as well as the nature of the primary particle; see Fig.\,2.
These
experiments have opened a new window in astronomy by extending the
photon spectrum to 20\,TeV, possibly beyond. Observations have
revealed spectacular TeV-emission from galactic supernova remnants
and nearby quasars, some of which emit most of their energy in very
short burst of TeV-photons.

But there is the dog that didn't bark. No evidence has emerged
for $\pi^0$ origin of the TeV radiation and, therefore, no cosmic
ray sources have yet been identified. Dedicated searches for photon beams from
suspected cosmic ray sources, such as the supernova remnants IC433
and $\gamma$-Cygni, came up empty handed. While not relevant to the
topic covered by this talk, supernova remnants are theorized to be the
sources of the bulk of the cosmic rays that are of galactic origin.
The evidence is still circumstantial.

The field of gamma ray astronomy is buzzing with activity to
construct second-generation instruments. Space-based detectors are
extending their reach from GeV to TeV energy with AMS and,
especially, GLAST, while the ground-based Cerenkov collaborations are
designing instruments with lower thresholds. In the not so far future
both techniques should generate overlapping measurements in the $10
{\sim} 10^2$~GeV energy range. All ground-based air Cerenkov experiments aim
at lower threshold, better angular- and energy-resolution, and a
longer duty cycle. One can however identify three pathways to reach
these goals:

\begin{enumerate}

\item
larger mirror area, exploiting the parasitic use of solar collectors
during nighttime (CELESTE, STACEY and\break SOLAR\,II),\cite{pare}
\item
better, or rather, ultimate imaging with the 17~m MAGIC mirror,\cite{magic}
\item
larger field of view using multiple telescopes (VERITAS, HEGRA and HESS).

\end{enumerate}

The Whipple telescope pioneered the atmospheric Cerenkov technique.
VERITAS\cite{veritas} is an array of 9 upgraded Whipple telescopes,€
each with a
field of view of 6 degrees. These can be operated in coincidence for
improved angular resolution, or be pointed at 9 different 6 degree
bins in the night sky, thus achieving a large field of view. The
HEGRA collaboration\cite{hegra} is already operating four telescopes
in coincidence and is building an upgraded facility with excellent
viewing and optimal location near the equator in Namibia.

There is a dark horse in this race: Milagro.\cite{milagro} The
Milagro idea is to lower the threshold of conventional air shower arrays to 100~GeV by
instrumenting a pond of five million gallons of ultra-pure water with
photomultipliers. For time-varying signals, such as bursts, the
threshold may be lower.

\subsection{High Energy Neutrino Telescopes}

Although neutrino telescopes have multiple interdisciplinary science
missions, the search for the sources of the highest-energy cosmic
rays stands out because it clearly identifies the size of the
detector required to do the science.\cite{gaisser}€ For guidance in estimating
expected signals, one makes use of data covering the highest-energy
cosmic rays in Fig.\,3 as well as known sources of non-thermal,
high-energy gamma
rays. Accelerating particles to TeV energy and above involves neutron
stars or black holes. As already explained in the context of Fig.\,5,
some fraction of them will interact in the radiation fields
surrounding the source, whatever it may be, to produce pions.€ These
interactions may also be hadronic collisions with ambient gas.€ In
either case, the neutral pions decay to photons while charged pions
include neutrinos among their decay products with spectra related to
the observed gamma-ray spectra. Estimates based on this relationship
show that a kilometer-scale detector is needed to see neutrino
signals.

The same conclusion is reached in specific models. Assuming, for instance, that gamma ray bursts are the cosmic
accelerators of the highest-energy cosmic rays, one can calculate
from textbook particle physics how many neutrinos are produced when
the particle beam coexists with the observed MeV energy photons in
the original fireball. We thus predict the observation of 10--100
neutrinos of PeV energy per year in a detector with a
kilometer-square effective area. In general, the potential scientific payoff of
doing neutrino astronomy arises from the great penetrating power of
neutrinos, which allows them to emerge from dense inner regions of
energetic sources.

Whereas the science is compelling, the real challenge has been to develop a reliable, expandable and affordable detector technology.
Suggestions to use a large volume of deep ocean water for high-energy neutrino
astronomy were made as early as the 1960s. In the case
of the muon neutrino, for instance, the neutrino ($\nu_\mu$)
interacts with a hydrogen or oxygen nucleus in the water and produces
a muon travelling in nearly the same direction as the neutrino. The
blue Cerenkov light emitted along the muon's $\sim$kilometer-long
trajectory is detected by strings of photomultiplier tubes deployed
deep below the surface. With the first observation of neutrinos in
the Lake Baikal and
the (under-ice) South Pole neutrino telescopes, there is optimism
that the technological challenges to build neutrino telescopes have
been met.

The first generation of neutrino telescopes, launched by the bold
decision of the DUMAND collaboration to construct such an
instrument, are designed to reach a large telescope area
and detection volume for a neutrino threshold of order 10~GeV. The
optical requirements of the detector medium are severe. A large
absorption length is required because it determines the spacings of
the optical sensors
and, to a significant extent, the cost of the detector. A long
scattering length is needed to preserve the geometry of the Cerenkov
pattern. Nature has been kind and offered ice and water as adequate
natural Cerenkov media. Their optical properties are, in fact,
complementary. Water and ice have similar attenuation length, with
the role of scattering and absorption reversed. Optics seems, at
present, to drive the evolution of ice and
water detectors in predictable directions: towards very large
telescope area in ice exploiting the long absorption length, and
towards lower threshold and good muon track reconstruction in water
exploiting the long scattering length.

DUMAND, the pioneering project located off the coast of Hawaii,
demonstrated that muons could be detected by this
technique, but the planned detector was never realized.
A detector composed of 96 photomultiplier tubes located deep in Lake
Baikal was the first to demonstrate the detection of neutrino-induced
muons in natural water.\cite{baikal} In the following years, {\it
NT-200} will be operated as a neutrino telescope with an effective
area between $10^3 {\sim} 5\times 10^3$\,m$^2$, depending on energy.
Presumably too small to detect neutrinos from extraterrestrial
sources, {\it NT-200} will serve as the prototype for a larger
telescope. For instance, with 2000 OMs, a threshold of€ $10 {\sim}
20$\,GeV and an effective area of $5\times10^4 {\sim} 10^5$\,m$^2$, an
expanded Baikal telescope would fill the gap between present
detectors and planned high-threshold detectors of cubic kilometer
size. Its key advantage would be low threshold.

The Baikal experiment represents a proof of concept for deep ocean
projects. These do however have the advantage of larger depth and optically
superior water. Their challenge is to find reliable and affordable
solutions to a variety of technological challenges for deploying a
deep underwater detector. The European collaborations
ANTARES\cite{antares} and NESTOR\cite{NESTOR} plan to deploy
large-area detectors in the Mediterranean Sea within the next year.
The NEMO Collaboration is conducting a site study for a
future kilometer-scale detector in the Mediterranean.\cite{NEMO}

The AMANDA collaboration, situated at the U.S. Amundsen-Scott South Pole
Station, has demonstrated the merits of natural ice as a Cerenkov
detector medium.\cite{B4} In 1996, AMANDA was able to
observe atmospheric neutrino candidates using only 80 eight-inch
photomultiplier tubes.\cite{B4}

\begin{figure}[t]
\centering\leavevmode
\epsfxsize=6.8cm
\epsffile{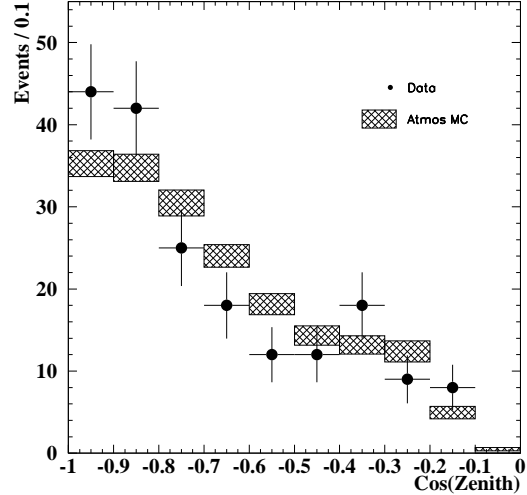}

\caption[]{Reconstructed zenith angle distribution. The
points mark the data and the shaded boxes a simulation of
atmospheric neutrino events, the widths of the boxes
indicating the error bars.
\label{fig:zenith}}
\end{figure}

With 302 optical modules instrumenting approximately 6000 tons of
ice, AMANDA extracted several hundred atmospheric neutrino events
from its first 130 days of data. AMANDA was thus the first
first-generation neutrino
telescope with an effective area in excess of 10,000 square meters
for TeV muons.\cite{nature00}  In rate and all characteristics the events are
consistent with atmospheric neutrino origin. Their energies are in
the 0.1--1\,TeV range. The shape of the zenith angle distribution is
compared to a simulation of the atmospheric neutrino signal in
Fig.~\ref{fig:zenith}. The variation of the measured rate with zenith
angle is reproduced by the simulation to within the statistical
uncertainty.€ Note that the tall geometry of the detector strongly
influences the dependence on zenith angle in favor of more vertical
muons.

\begin{figure}[t]
\centering\leavevmode
\epsfxsize=6.8cm
\epsffile{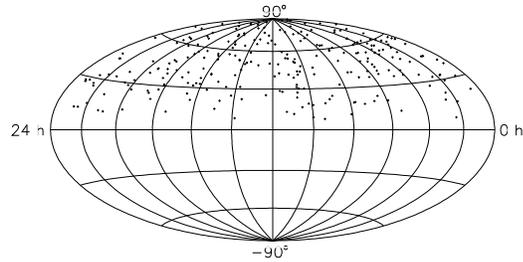}

\caption[]{Distribution in declination and right ascension of the up-going
events on the sky. \label{fig:skyplot}}
\end{figure}


The arrival directions of the neutrinos are shown in
Fig.~\ref{fig:skyplot}. A statistical analysis indicates no evidence
for point sources in this sample.€ An estimate of the energies of the
up-going muons (based on simulations of the number of reporting
optical modules) indicates that all events have energies consistent
with an atmospheric neutrino origin.€ This enables AMANDA to reach a
level of sensitivity to a diffuse flux of high energy
extra-terrestrial neutrinos of order\cite{nature00} $dN/dE_{\nu} = 10^{-6}
E_{\nu}^{-2} \rm\, cm^{-2}\, s^{-1}\,sr^{-1}\,€€ GeV^{-1}, $ assuming
an€ $E^{-2}$ spectrum. At this level they exclude a variety of
theoretical models which assume the hadronic origin of TeV photons
from active galaxies and blazars.\cite{stecker} Searches for
neutrinos from gamma-ray bursts, for magnetic monopoles, and for a
cold dark matter signal from the center of the Earth are also in
progress€ and, with only 138 days of data, yield limits comparable to
or better than those from smaller underground neutrino detectors that
have operated for a much longer period.

In January 2000, AMANDA-II was completed. It consists of 19 strings
with a total of 677 OMs arranged in concentric circles, with the ten
strings from AMANDA forming the central core of the new detector.
First data with the expanded detector indicate an atmospheric
neutrino rate increased by a factor of three, to 4--5 events per day.
AMANDA-II has met the key challenge of neutrino astronomy: it has
developed a reliable, expandable, and affordable technology for
deploying a kilometer-scale neutrino detector named IceCube.

\begin{figure}[h]
\epsfxsize=6.6cm
\centering\leavevmode
\epsffile{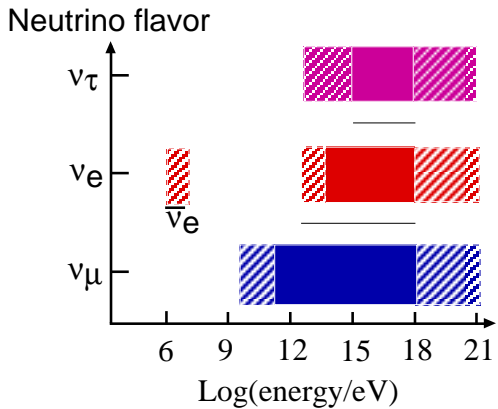}
\caption{Although IceCube detects neutrinos of any flavor above a threshold of $\sim 0.1$\,TeV, it can identify their flavor and measure their energy in the ranges shown. Filled areas: particle identification, energy, and angle. Shaded areas: energy and angle.}
\end{figure}

\begin{figure}[t]
\epsfxsize=6.8cm
\epsffile{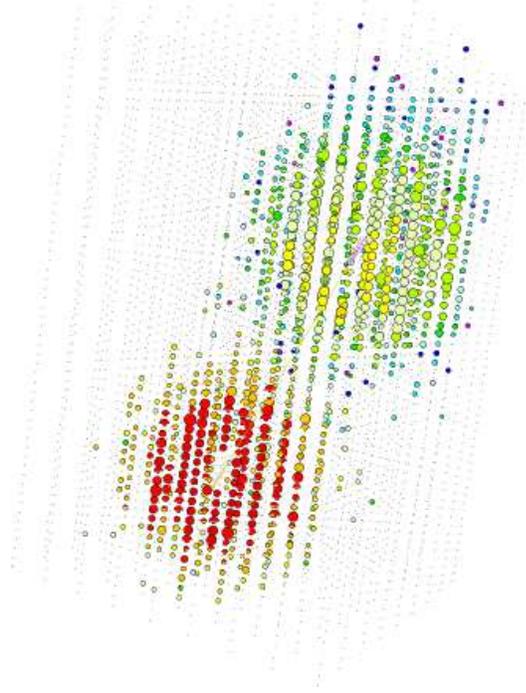}
\caption{Simulation of a ultra-high energy tau-lepton by the interaction of a 10 million GeV tau-neutrino, followed by the decay of the secondary tau-lepton. The color represents the time sequence of the hits (red-orange-yellow-green-blue). The size of the dots corresponds to the number of photons detected by the individual photomultipliers.}
\end{figure}

IceCube is an instrument optimised to detect and characterize sub-TeV
to multi-PeV neutrinos of all flavors (see Fig.\,8) from
extraterrestrial sources.€ It will consist of 80 strings, each with 60 10-inch photomultipliers spaced 17~m apart.
The deepest module is 2.4~km below the surface.€ The strings are
arranged at the apexes of equilateral triangles 125~m on a side. The
effective detector volume is about a cubic kilometer, its precise
value depending on the characteristics of the signal. IceCube will
offer great advantages
over AMANDA II beyond its larger size: it will have a much higher
efficiency to reconstruct tracks, map showers from electron- and
tau-neutrinos (events where both the production and decay of a $\tau$
produced by a $\nu_{\tau}$ can be identified; see Fig.\,9) and, most
importantly, measure neutrino energy. Simulations indicate that the
direction of muons can be determined with sub-degree accuracy and
their energy measured to better than 30\% in the logarithm of the
energy. Even the direction of showers can be reconstructed to better
than 10$^\circ$ in both $\theta$, $\phi$ above 10\,TeV. Simulations
predict a linear response in energy of better than 20\%. This has to
be contrasted with the logarithmic energy resolution of
first-generation detectors. Energy resolution is critical because,
once one establishes that the energy exceeds 100~TeV, there is no atmospheric neutrino background in a kilometer-square detector.

At this point in time, several of the new instruments, such as the partially deployed Auger array and HiRes to Magic to Milagro and AMANDA~II, are less than one year from delivering results. With rapidly growing observational capabilities, one can express the realistic hope that the cosmic ray puzzle will be solved soon. The solution will almost certainly reveal unexpected astrophysics, if not particle physics.

\section*{Acknowledgements}
I thank Concha Gonzalez-Garcia and Vernon Barger for comments on the 
manuscript. This research was supported in part by the U.S.~Department of Energy
under Grant No.~DE-FG02-95ER40896 and in part by the University of
Wisconsin Research Committee with funds granted by the Wisconsin Alumni
Research Foundation.


\begin{thebibliography}{99}
\frenchspacing
%
\bibitem{flyes} D. J. Bird {\it et al.}, {\it Phys. Rev. Lett.} {\bf
71}, 3401 (1993).
%
\bibitem{yakutsk} N. N. Efimov {\it et al.}, {\it ICRR Symposium on
Astrophysical Aspects of the Most Energetic Cosmic Rays}, ed. M.~Nagano
and F. Takahara (World Scientific, 1991).
%
\bibitem{web}
http://\\
www.hep.net/experiments/all\_sites.html, provides information
on experiments discussed in this review. For a few exceptions, I will
give separate references to articles or websites.

%
\bibitem{WatsonZas}
M.~Ave {\it et al.}, {\it Phys. Rev. Lett.} {\bf 85}, 2244 (2000).
%
\bibitem{agasa}
http://\\
www-akeno.icrr.u-tokyo.ac.jp/AGASA/
%
\bibitem{ICRC}
Proceedings of the International Cosmic Ray Conference, Hamburg,
Germany, August 2001. Some of the results described here can be
found in the rapporteur's talks of this meeting which was held two
weeks after this conference.
%
\bibitem{vazquez}
R.~A.~Vazquez {\it et al.}, {\it Astroparticle Physics} {\bf 3}, 151 (1995).

\bibitem{stecker2}
F.~W.~Stecker and M.~H.~Salamon, astro-ph/9808110 and references therein.

\bibitem{han}
J.\,Alvarez-Muniz {\it et al.}, hep-ph/0107057; R.~Emparan {\it et 
al.}, hep-ph/0109287 and references therein.


\bibitem{weekes}
T.~C.~Weekes, Status of VHE Astronomy c.2000, {\it Proceedings of the
International Symposium on High Energy Gamma-Ray Astronomy},
Heidelberg, June 2000, astro-ph/0010431;
R.~A.~Ong, {\it XIX International Symposium on Lepton and Photon
Interactions at High Energies}, Stanford, August 1999, hep-ex/0003014.

\bibitem{pare}
E.~Pare {\it et al.}, astro-ph/0107301.

\bibitem{magic}
J. Cortina for the MAGIC collaboration, {\it Proceedings of the Very High
Energy Phenomena in the Universe}, Les Arcs, France, January 20--27,
2001, astro-ph/0103393.

\bibitem{veritas}
http://veritas.sao.arizona.edu/

\bibitem{hegra}
http://hegra1.mppmu.mpg.de

\bibitem{milagro}
http://\\
www.igpp.lanl.gov/ASTmilagro.html

\bibitem{gaisser}
For reviews, see T.K.~Gaisser, F.~Halzen and T.~Stanev, {\it Phys. Rep.} {\bf
258}(3), 173 (1995); J.G.~Learned and K.~Mannheim, {\it Ann. Rev. Nucl.
Part. Science} {\bf 50}, 679 (2000); R.~Ghandi, E.~Waxman and
T.~Weiler, review talks at Neutrino 2000, Sudbury, Canada (2000).

\bibitem{baikal}
I. A. Belolaptikov {\it et al.}, {\it Astroparticle Physics} {\bf 7}, 263 (1997).

\bibitem{antares}
E. Aslanides {\it et al}, astro-ph/9907432 (1999).

\bibitem{NESTOR} L. Trascatti, in
{\it Procs. of the 5th International Workshop on ``Topics in
Astroparticle and Underground Physics (TAUP\,97)}, Gran Sasso, Italy,
1997, ed. by A. Bottino, A.~di\,Credico, and P.~Monacelli, {\it Nucl.
Phys.} {\bf B70} (Proc.
Suppl.), p.\,442 (1998).


\bibitem{NEMO} Talk given at the
{\it International Workshop on Next Generation Nucleon
Decay and Neutrino Detector (NNN\,99)},  Stony Brook, 1999, Proceedings
to be published by AIP.

\bibitem{B4}
The AMANDA collaboration, {\it Astroparticle Physics}, {\bf 13}, 1 (2000).

\bibitem{nature00}
E. Andres {\it et al.}, {\it Nature} {\bf 410}, 441 (2001).

\bibitem{stecker}
F. Stecker, C. Done, M. Salamon, and P. Sommers, {\it Phys. Rev. Lett.} {\bf
66}, 2697 (1991); erratum {\it Phys. Rev. Lett.} {\bf 69}, 2738 (1992).

\end{thebibliography}
\end{document}